\documentclass[lettersize,journal]{IEEEtran}
\usepackage{tikz}
\usepackage{amsmath}
\usepackage{cite}
\usepackage{amsmath,amssymb,amsfonts}
\usepackage{algorithmic}
\usepackage{graphicx}
\usepackage{booktabs}
\usepackage{textcomp}
\usepackage{multirow}
\usepackage{xcolor}
\usepackage{colortbl}
\usepackage{tabularx}
\usepackage{caption}
\usepackage[normalem]{ulem}
\usepackage{lipsum}
\usepackage{enumitem}
\usepackage{multirow}
\usepackage{makecell}
\usepackage{pifont}
\usepackage[compact]{titlesec}
\usepackage{siunitx}
\usepackage{makecell}
\usepackage{wrapfig}
\usepackage{tablefootnote}
\useunder{\uline}{\ul}{}
\def\BibTeX{{\rm B\kern-.05em{\sc i\kern-.025em b}\kern-.08em
    T\kern-.1667em\lower.7ex\hbox{E}\kern-.125emX}}

\usepackage{tabularray}
\usepackage{tcolorbox}

\newtcolorbox{takeawaybox}{
  colback=gray!10!white,
  colframe=gray!90!black,
  title=Takeaways,
  sharp corners,
  top=2mm,
  bottom=2mm,
  left=2mm,
  right=2mm,
  boxrule=0.5mm
}

\newtcolorbox{openquestionbox}{
  colback=gray!10!white,
  colframe=gray!90!black,
  title=Open questions,
  sharp corners,
  top=0.1mm,
  bottom=0.1mm, 
  left=0.1mm,
  right=0.1mm,
  boxrule=0.1mm
}

\usepackage{amsmath,amsfonts}
\usepackage{algorithmic}
\usepackage{array}
\usepackage[caption=false,font=normalsize,labelfont=sf,textfont=sf]{subfig}
\usepackage{textcomp}
\usepackage{stfloats}
\usepackage{url}
\usepackage{verbatim}
\usepackage{graphicx}
\def\BibTeX{{\rm B\kern-.05em{\sc i\kern-.025em b}\kern-.08em
    T\kern-.1667em\lower.7ex\hbox{E}\kern-.125emX}}
\usepackage{balance}
\usepackage{tikz}
\usepackage{hyperref} 

\begin{document}
\title{Breaking TinyML: Why Quantized Neural Networks Need Domain-Specific Security Analysis}
\author{\IEEEauthorblockN{Jacob Huckelberry\IEEEauthorrefmark{6}\IEEEauthorrefmark{1}\IEEEauthorrefmark{2}, Andrea Mattia Garavagno\IEEEauthorrefmark{6}\IEEEauthorrefmark{1}\IEEEauthorrefmark{3}\IEEEauthorrefmark{4}, \\ Yuke Zhang\IEEEauthorrefmark{5}, Peter A. Beerel\IEEEauthorrefmark{5}, James Mickens\IEEEauthorrefmark{1}, Vijay Janapa Reddi\IEEEauthorrefmark{1}} \\
\IEEEauthorblockA{\IEEEauthorrefmark{1}Harvard University, }
\IEEEauthorblockA{\IEEEauthorrefmark{2}Draper Laboratory, }
\IEEEauthorblockA{\IEEEauthorrefmark{3}University of Genoa, }
\IEEEauthorblockA{\IEEEauthorrefmark{4}Scuola Superiore Sant'Anna, } \\
\IEEEauthorblockA{\IEEEauthorrefmark{5}University of Southern California} \\
\IEEEauthorblockA{\IEEEauthorrefmark{6}These authors contributed equally to this work and share first authorship}
}

\maketitle

\begin{tikzpicture}[remember picture,overlay]
\node[anchor=south,yshift=15pt] at (current page.south) {
    \parbox{\textwidth}{
        \centering \scriptsize
        This work has been accepted for publication in IEEE Micro. \copyright~2026 IEEE. Personal use of this material is permitted.  Permission from IEEE must be obtained for all other uses, in any current or future media, including reprinting/republishing this material for advertising or promotional purposes, creating new collective works, for resale or redistribution to servers or lists, or reuse of any copyrighted component of this work in other works. Final publication is available at \url{https://doi.org/10.1109/MM.2026.3666128}.
    }
};
\end{tikzpicture}

\begin{abstract}
Most TinyML hardware focus on supporting Quantized Neural Networks (QNNs) to meet stringent constraints on power consumption, size, and cost. Despite this, the security aspects of quantization within TinyML hardware remain largely unexplored. Although previous studies indicate that QNNs demonstrate similar or enhanced robustness when compared to full-precision Deep Neural Networks (DNNs) against typical evasion attacks, no attack strategies tailored specifically for TinyML hardware have been proposed yet. This paper addresses the aforementioned shortfall by demonstrating how a two-step attack pipeline can surpass the current state-of-the-art in the QNN context and shows the need for more hardware-aware security research.
\end{abstract}


\section{Introduction}
\label{sec:intro}
The rapid proliferation of Tiny Machine Learning (TinyML) has introduced unique security challenges that fundamentally differ from traditional machine learning deployments. TinyML mandates quantization to meet the stringent resource constraints of commodity electronics. Furthermore, most TinyML hardware accelerators, including Google's Coral Edge TPU, Qualcomm's Hexagon NPU, and NE16, exclusively support Quantized Neural Networks (QNNs) \cite{seshadri2022evaluation, conti2018xnor}, creating a distinct computational paradigm that necessitates specialized security analysis for hardware-accelerated AI systems.

Despite the widespread adoption of quantization in TinyML hardware, the security implications of this hardware-software co-design choice remain underexplored. Current understanding of QNN security relies heavily on applying state-of-the-art Deep Neural Network (DNN) attack methodologies to quantized models. Preliminary studies \cite{thorsteinsson2024adversarial} have crafted adversarial examples for QNNs using model evasion attacks originally designed for DNNs. These studies revealed that QNNs show no loss in model robustness and, in some cases, demonstrate better robustness than their full-precision counterparts \cite{costa2024david}.

However, a critical gap exists in the current security landscape: no attack specifically designed to exploit the hardware characteristics of QNN accelerators has been proposed \cite{costa2024david}. The unique properties of quantized hardware implementations, including discrete value spaces, gradient discontinuities, and specialized computational patterns, create potential attack vectors that conventional DNN attacks cannot exploit, yet this hardware-specific attack surface remains unexplored.

The practical implications are significant. TinyML devices are deployed at massive scale (smart doorbells, parking monitors, industrial sensors) and firmware updates are difficult or impossible for many embedded systems. An attacker who extracts a model once can then craft \textit{reusable} physical adversarial perturbations: a sticker on a license plate that makes a car invisible to parking fee detectors, or a pattern printed on clothing that defeats person detection in security cameras. Unlike digital attacks requiring per-image computation, these physical perturbations work persistently whenever the deployed model encounters them, amplifying the impact across millions of devices.

To address this critical gap, we apply for the first time a well-studied two-step attack pipeline for DNNs \cite{papernot2016transferability, juuti2019prada} to circumvent the typical gradient discontinuities preventing the application of white-box attacks in the QNN domain. Our approach achieves up to 47.30\% accuracy reduction on CIFAR-10 (reducing model accuracy from 85.30\% to 38.00\%, compared to only 19.70\% for the best gray-box attack) while requiring only 50,000 total queries. In contrast, black-box attacks like Boundary Attack and GeoDA require thousands of queries \textit{per adversarial example}; generating adversarial examples for 1,000 test images would require millions of total queries, making our approach orders of magnitude more efficient.

Moreover, this attack effectiveness is hardware-specific: when applied to full-precision models, our methodology underperforms compared to existing attacks. This performance differential challenges the assumption that quantization inherently improves model robustness. Instead, our findings suggest that conventional attacks may simply be ill-suited for the quantized hardware domain, creating an artificial impression of enhanced security.

The contributions of this work are fourfold:
\begin{enumerate}
\item We demonstrate the first successful circumvention of hardware-imposed gradient discontinuities in QNNs through surrogate-based white-box attacks achieving 82-84\% fidelity.
\item We show, on CIFAR-10, superior performance against quantized hardware models compared to state-of-the-art gray-box attacks (over 20\% better accuracy reduction) successfully overcoming gradient-masking, while maintaining practical query efficiency.
\item We benchmark our approach on commercial TinyML hardware and demonstrate the fastest interaction time compared to the most query-efficient state-of-the-art alternative.
\item We prove that the attack's effectiveness is hardware-specific, thereby establishing the critical need for domain-specific security research in TinyML systems.
\end{enumerate}

These findings have significant implications for the rapidly expanding TinyML hardware ecosystem, where resource constraints often preclude the implementation of comprehensive security measures, making hardware-aware attack methodologies and corresponding defenses increasingly important.

\section{Related Works}
\label{sec:related_works}
The intersection of DNN quantization and adversarial robustness has emerged as an important research area, though current understanding remains limited by the application of conventional attack methodologies to quantized models deployed on TinyML hardware.

\subsection{Quantization and Adversarial Robustness}
Early investigations into the security implications of quantization revealed counterintuitive findings regarding model robustness. Previous studies demonstrated that quantization of neural networks does not inherently lead to loss in model robustness and can introduce resilience against the transferability of adversarial attacks \cite{thorsteinsson2024adversarial}. Building on these preliminary findings, recent comprehensive work by Costa et al. \cite{costa2024david} found that 8-bit QNNs deployed on edge hardware are more robust to some state-of-the-art model evasion attacks than their full-precision counterparts. Their study systematically evaluated both gray-box attacks, which can query the model an unlimited number of times while obtaining complete output information in the form of confidence scores, and black-box attacks, which can only access the predicted class labels. However, their evaluation was limited to directly applying existing DNN attacks without considering hardware-specific vulnerabilities.

\subsection{State-of-the-Art Attack Methodologies}
The evaluation of QNN security has relied on adapting existing DNN attack frameworks to the quantized domain. Among the state-of-the-art gray-box attacks applied to QNNs on TinyML hardware, the Zeroth Order Optimization (ZOO) attack and the Square Attack (SA) have shown varying degrees of effectiveness. The ZOO attack leverages zeroth order stochastic coordinate descent along with dimension reduction, hierarchical attack, and importance sampling techniques to craft adversarial perturbations for given input images. In contrast, the SA attack employs random search to craft squared perturbations capable of fooling the classifier for a given input. While effective against DNNs, our results show these methods underperform when targeting quantized hardware due to gradient masking effects.

For black-box attack evaluation, researchers have applied the Boundary Attack (BA) and the Geometric Decision-based Attack (GeoDA) to QNNs. The BA starts from a large adversarial perturbation and then seeks to reduce the perturbation while maintaining its adversarial properties, whereas GeoDA exploits the low mean curvature of the decision boundary in the vicinity of data samples to effectively estimate the normal vector to the decision boundary and produce minimal perturbations to cross it. While these achieve high success rates, they require prohibitive query counts unsuitable for practical hardware or TinyML deployment scenarios.

\subsection{Limitations of Current Approaches}
An important characteristic distinguishing QNN security evaluation from traditional DNN analysis is the absence of white-box attack assessments. Black-box attacks can require large amounts of queries to craft adversarial examples, while gray-box attacks can be more efficient in this aspect by leveraging confidence scores of the attacked DNN.

The barrier to white-box analysis in QNNs is the gradient discontinuity problem. Quantization operations map continuous inputs to discrete output levels, creating a staircase activation function.

In this environment, exact gradient calculation fails because:
\begin{itemize}
    \item Zero Gradients: On the flat sections of the 'steps,' the derivative is exactly zero, providing no directional information for the attack algorithm to update its noise;
    \item Undefined Gradients: At the vertical jumps between steps, the derivative is undefined.
\end{itemize}
    
This limitation represents a significant gap in current QNN security analysis, as white-box attacks typically achieve higher success rates than their gray-box and black-box counterparts in the full-precision domain. The inability to compute exact gradients has effectively eliminated an entire class of potent attack methodologies from QNN security evaluation on TinyML hardware. Our work circumvents this limitation through surrogate model construction.

\subsection{Research Gap and Motivation}
The current state of QNN security research reveals a fundamental limitation: all existing approaches apply attack methodologies originally designed for full-precision DNNs to quantized models running on specialized hardware. This approach assumes that the security characteristics of hardware-accelerated QNNs can be adequately understood through the lens of conventional attacks, potentially overlooking vulnerabilities specific to the quantized hardware domain.

Our work addresses this gap by proposing for the first time the usage of a well-studied two-step attack pipeline for DNNs \cite{papernot2016transferability, juuti2019prada}, in the QNN hardware domain, enabling the usage of white-box attacks through surrogate model construction. While surrogate-based model extraction has been explored extensively in the DNN domain \cite{orekondy2019knockoff, correia2018copycat}, no prior work has applied this pipeline specifically to circumvent gradient discontinuities in hardware-deployed QNNs. In the first phase, a surrogate model is built using extraction attacks, then white-box attacks are used to craft adversarial examples leveraging exact gradient calculations from the surrogate. Finally the crafted adversarial examples are transferred to the target QNN running on TinyML hardware.

The latter methodology overcomes state-of-the-art gray-box attacks on MLPerf Tiny models \cite{banbury2021mlperf}, a TinyML hardware benchmark, while providing a more practical and scalable alternative to direct black-box attacks thanks to a one-time query performed during the model extraction phase. Importantly, the performance of the adopted methodology falls behind in the full-precision context, highlighting the need for hardware-aware attacks and challenging assumptions about the inherent robustness advantages of quantized models on accelerators.

\section{Threat Model}
\label{sec:threat_model}
We define a threat model outlining the adversary's objectives, capabilities, and knowledge, providing a practical framework for evaluating attacks against TinyML hardware.

\subsection{Adversary Goals}
The adversary's primary objective is to compromise the integrity of the targeted quantized neural network by slightly perturbing benign inputs causing the model to misclassify them. While there are multiple pathways to achieve model compromise \cite{papernot2016transferability, juuti2019prada}, this work focuses specifically on a two-step attack sequence that systematically combines model extraction and evasion attacks to exploit vulnerabilities unique to the quantized hardware domain.

\subsection{Adversary Capabilities}
The attacker has temporary physical access to the deployed TinyML device, which is sufficient to perform queries and inject crafted inputs during the model extraction
phase, but not necessarily to extract firmware or permanently modify it. We assume a Man-in-the-Middle (MitM) or direct injection setup where the adversary can bypass the physical sensor and inject digital signals directly into the hardware interface (e.g., SPI, I2C, or MIPI CSI). Furthermore, regarding output observation, the attacker is restricted to observing external device behaviors. This includes monitoring GPIO state transitions (e.g., a 'detect' pin going high), analyzing power consumption traces, or observing network side-channels (e.g., WiFi packet timing).

\subsection{Adversary Knowledge}
To make this analysis as realistic as possible while remaining practically relevant, we assume the attacker operates under a black-box threat model with several important constraints and assumptions that align with real-world TinyML hardware deployment scenarios. The attacker does not have access to the model architecture details, trained weights, or gradient information. They may only interact with the model by feeding it inputs and observing the corresponding outputs, simulating the typical interaction pattern available to external adversaries targeting deployed TinyML accelerators.

We assume a set of unlabeled samples drawn from the Original Domain Data (ODD) distribution is known to the attacker for a one-time model extraction phase. This reflects a realistic scenario where an attacker targeting a specific application, such as a commercial person-detection doorbell, can collect imagery from well-known TinyML person detection datasets.

The attacker does know that the target model employs 8-bit integer quantization. This assumption is reasonable given that quantization is a well-known characteristic of TinyML hardware deployments, often documented in accelerator specifications or inferable from hardware constraints.

Additionally, we assume the model only outputs hard labels rather than the full softmax probability distribution. This constraint aligns well with TinyML hardware deployment practices that require minimization of computational overhead and information leakage. Many TinyML accelerators implement this restriction as both a performance optimization and a security measure, making our assumption highly realistic for practical hardware deployments.

\section{TinyML Model Extraction and Evasion}
\label{sec:proposal}
\begin{figure}[t]
    \centering
    \includegraphics[width=\linewidth]
    {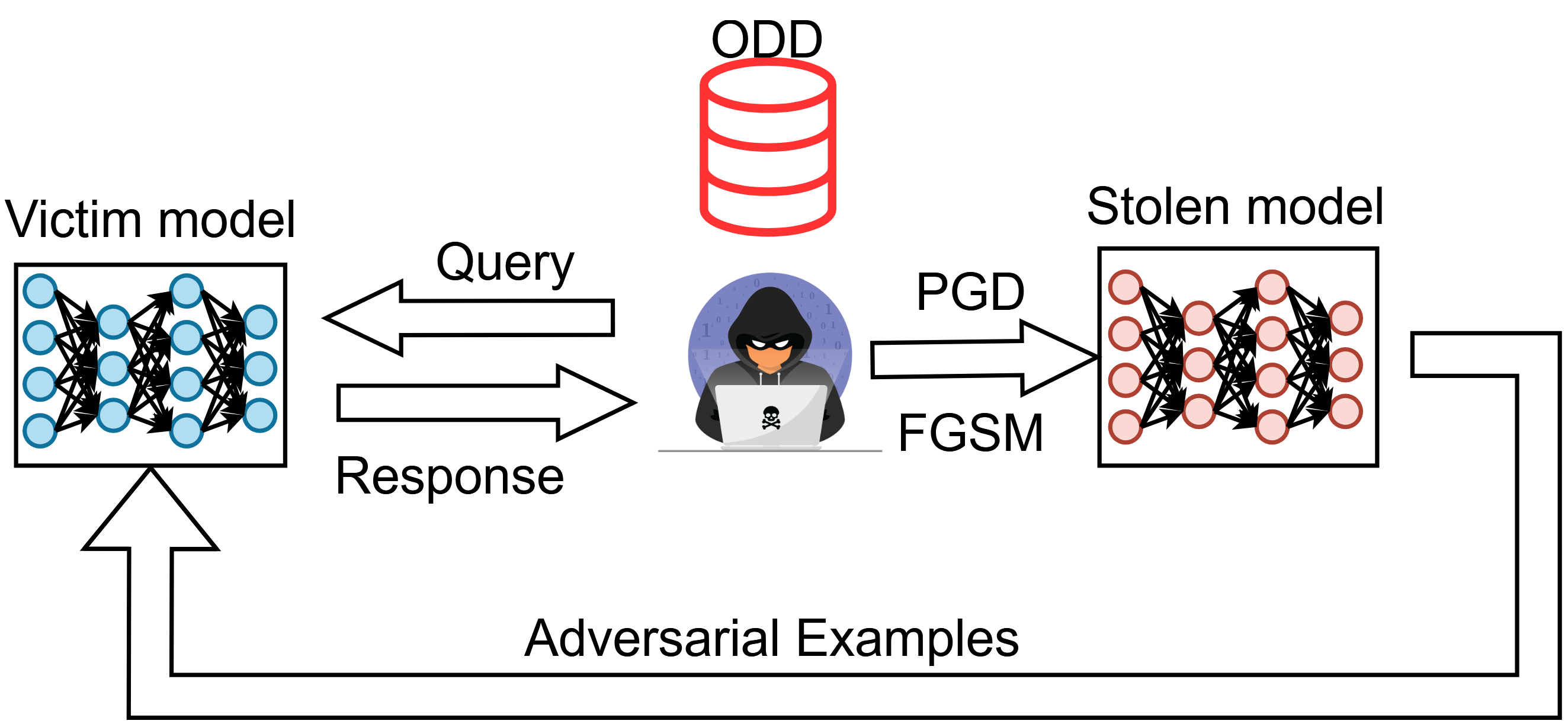}
    \caption{A figure describing the two-step attack. First, the attacker performs a model extraction attack using Copycat CNN. For this attack, the adversary uses the ODD to query the target model, and uses the input output pairs to train a surrogate model. The attacker then uses the surrogate model to craft adversarial examples using PGD and FGSM. These adversarial examples are then transferred to the target model.}
    \label{fig:attack}
\end{figure}

The attack pipeline operates by first approximating the target model through a model extraction attack, then leveraging the resulting surrogate model to craft white-box adversarial examples using established methods such as the Fast Gradient Signed Method (FGSM) \cite{goodfellow2014explaining} and Projected Gradient Descent (PGD) \cite{madry2017towards}. These adversarial examples are subsequently transferred to the target model running on TinyML hardware to evaluate their effectiveness. A comprehensive visual depiction of this attack sequence is presented in Figure \ref{fig:attack}.

\subsection{Model Extraction Step}
The foundation of our attack pipeline lies in conducting a strategic model extraction attack to create a surrogate model that strongly approximates the decision boundaries and behavior of the target quantized model deployed on hardware. While several approaches exist for model extraction, we selected Copycat CNN \cite{correia2018copycat} based on its compatibility with our defined threat model and its minimal operational requirements.

Copycat CNN is particularly well-suited to our threat model because it operates under minimal assumptions that align with realistic TinyML hardware deployment constraints. The method requires only that the target model outputs hard labels, the attacker has access to unlabeled data from an appropriate distribution, and that the attacker has selected a suitable surrogate model architecture for the extraction process. 

The Copycat CNN extraction process utilizes the ODD to systematically query the target model and collect the input-output pairs necessary for surrogate training. ODD refers to the dataset on which the model was originally trained and validated, providing the most direct approximation of the target's training distribution. The collected input-output pairs are then used to train a surrogate model.

Importantly, the surrogate model need not perfectly replicate the target's outputs; it only needs to approximate the target's decision boundary geometry. A surrogate achieving 82-84\% fidelity (agreement on roughly 5 out of 6 predictions) is sufficient for adversarial transfer because adversarial examples exploit the structure of decision boundaries rather than exact weight values. Perturbations that cross boundaries in the surrogate will likely cross similar boundaries in the target, enabling effective attacks despite imperfect extraction.

\subsection{Model Evasion Step}
Once the attacker has successfully crafted a sufficiently accurate surrogate model, or ensemble of surrogate models, they can proceed to the adversarial example generation phase. The key advantage of this approach is that the attacker now has white-box access to the surrogate model(s), enabling the use of gradient-based white-box evasion attacks to generate adversarial examples that can be transferred to the target model on the hardware accelerator.

For the evasion component of our attack pipeline, we selected FGSM \cite{goodfellow2014explaining} and PGD \cite{madry2017towards} due to their demonstrated effectiveness and strong transferability properties across different model architectures and hardware domains. Both attacks represent well-established methods in the adversarial machine learning literature and provide complementary approaches to adversarial example generation.

Both FGSM and PGD generate perturbations constrained by the $L_\infty$ norm, creating perturbations that are dense and uniform across the target image. This perturbation characteristic is particularly important for transferability, as the uniform nature of the modifications means they are applied broadly across the input space and do not rely on specific input features that may be unique to the surrogate model architecture.

\subsection{Benchmark Selection and Target Models}
For our experimental evaluation, we selected the benchmark from the TinyML industry-standard MLPerf Tiny suite \cite{banbury2021mlperf}, which provides models and datasets specifically designed for hardware accelerator deployments. We target two key models representative of real-world hardware constraints: a customized ResNetv1 for CIFAR-10, and a MobilenetV1-0.25 for Visual Wake Words (VWW). These models are optimized for different hardware accelerator architectures and application domains within the TinyML ecosystem.

\subsection{Model Extraction Implementation}
In the model extraction phase, we use Copycat CNN \cite{correia2018copycat} as implemented in the Adversarial Robustness Toolbox (ART) to generate surrogate models that approximate the quantized behavior of models running on TinyML hardware. The selection of surrogate architectures balances approximation quality with the computational constraints typical of attacking deployed hardware systems.

For the CIFAR-10 hardware reference model, we employ ResNet50 as the surrogate, while for the VWW hardware model, we use MobileNetV1. Both utilize ImageNet pre-trained weights from Keras Applications, providing a foundation for capturing hardware-specific behaviors.

We systematically query the target hardware with 50,000 samples from the ODD training set. The resulting input-output pairs fine-tune the surrogate models for 10 epochs, balancing extraction quality with the query limitations imposed by real hardware deployments.

\subsection{Hardware Implementation}
A microcontroller, programmed to emulate the sensor's interface, injects samples to the target TinyML hardware and then monitors the GPIO activity to detect the predicted label, according to the threat model proposed in section \ref{sec:threat_model}. The microcontroller communicates via SPI with a host computer and is used both to query the target TinyML hardware in the model extraction phase and to test the adversarial examples crafted by the adopted attack.

\subsection{Hardware Selection}
We use three commercial TinyML hardware platforms to evaluate the cost of the model extraction phase, namely: the NUCLEO-F401RE, the NUCLEO-G474RE, and the NUCLEO-H743ZI2. An OV7670 image sensor is connected to each board. I2C interface and DMA are set on every board to automatically transfer the data acquired from the sensor to the memory buffer used as input of the programmed neural network without impacting the computational power available to the TinyML hardware. The OV7670 outputs a maximum of 30 frame per second.

\subsection{Adversarial Example Generation and Transfer}
Following surrogate model creation, we generate adversarial examples using both PGD and FGSM attacks. We craft adversarial examples using 1,000 test samples with $\epsilon = 8/255$, the standard for evaluating hardware robustness \cite{croce2020robustbench}. The $\epsilon$ parameter defines the limit on how much the attacker is allowed to change the original image to fool the model, ensuring the attack remains "imperceptible" to the human eye. These examples are then transferred to target models running on both int-8 quantized hardware (actual deployment) and full-precision implementations to assess hardware-specific vulnerabilities.

\section{Evaluation}
\label{sec:results}
We evaluate our hardware-aware attack pipeline on two MLPerf Tiny benchmarks, demonstrating that our two-step approach effectively exploits vulnerabilities specific to quantized models deployed on TinyML hardware. Our comprehensive analysis examines each attack phase across both benchmarks and compares performance against state-of-the-art methods.

\subsection{Evaluation Methodology}
We assess our attack pipeline along four key dimensions for each benchmark dataset (CIFAR-10 in Section \ref{sec:cifar10_results} and VWW in Section \ref{sec:vww_results}):

\textbf{Model Extraction Fidelity} measures how accurately our surrogate models replicate the behavior of hardware-deployed targets. High fidelity, where the surrogate mirrors the hardware model's decision boundaries, is essential for generating effective adversarial examples against accelerators.

\textbf{Adversarial Transferability} evaluates whether adversarial examples crafted on surrogate models successfully fool both int-8 quantized hardware and full-precision models. This reveals whether our attack exploits hardware-specific vulnerabilities or general model weaknesses.

\textbf{Comparison with State-of-the-Art} benchmarks our approach against existing gray-box and black-box attacks, demonstrating superior effectiveness and query efficiency critical for practical hardware deployments.

\textbf{Feasibility on TinyML Hardware} measures the interaction time with the target TinyML hardware required by the proposed attack.

Throughout our evaluation, we compare attack performance between int-8 quantized models (actual hardware deployment) and full-precision counterparts. This comparison confirms that our approach specifically targets hardware-induced vulnerabilities, as detailed in Section~\ref{sec:discussion}.

\subsection{CIFAR-10 Results}
\label{sec:cifar10_results}

Our attack pipeline achieves significant accuracy reduction on the CIFAR-10 hardware benchmark, with PGD reducing model accuracy from 85.30\% to 38.00\% (47.30\% reduction) and FGSM achieving 35.30\% reduction on int-8 quantized models. This attack effectiveness is domain-specific: when applied to full-precision models, our approach underperforms compared to state-of-the-art attacks, highlighting that we are exploiting hardware-specific vulnerabilities rather than general model weaknesses. Furthermore, our approach requires only 50,000 queries total, making it orders of magnitude more practical than black-box alternatives that can require millions of queries. Next, we now examine each phase of the attack to understand how this effectiveness is achieved.

\subsubsection{Phase 1: Model Extraction Fidelity}
\label{sec:cifar10_extraction}

\begin{table}[!t]  
  \centering
  \caption{Copycat CNN results on the target ResNetV1 using the ODD dataset CIFAR-10 (upper section of the table) and on target MobileNetV1 with the alpha parameter set at 0.25 using the ODD dataset VWW (lower section of the table).}
  \label{tab:extract_odd}
  \rowcolors{2}{gray!20}{white}  
  \begin{tabular}{@{}lllll@{}}  
    \toprule
    \textbf{Target Model} & \textbf{Surr. Arch.} & \textbf{Fidelity} & \textbf{Target Acc.} & \textbf{Surr. Acc.} \\ 
    \midrule
    \rowcolor{white}
    \multicolumn{5}{c}{CIFAR-10} \\
    ResNetV1 (Int-8) & ResNet-50 & 82.70\% & 84.30\% & 83.00\% \\
    ResNetV1 (FP-32) & ResNet-50 & 83.00\% & 84.30\% & 82.30\% \\
    ResNetV1 (Int-8) & MbNetV1 & 83.80\% & 84.30\% & 84.40\% \\
    ResNetV1 (FP-32) & MbNetV1 & 83.40\% & 84.30\%  & 84.80\% \\
    \midrule
    \rowcolor{white}
    \multicolumn{5}{c}{Visual Wake Words} \\
    MbNet025 (Int-8) & ResNet-50 & 79.40\% & 85.60\% & 69.60\% \\
    MbNet025 (FP-32) & ResNet-50 & 81.40\% & 85.60\% & 71.9\% \\
    MbNet025 (Int-8) & MbNetV1 & 82.10\% & 85.60\% & 72.00\% \\
    MbNet025 (FP-32) & MbNetV1 & 80.80\% & 85.60\%  & 70.60\% \\    
    \bottomrule
  \end{tabular}
\end{table}

The basis of our attack's success lies in achieving high fidelity during model extraction. As shown in Table \ref{tab:extract_odd}, our extracted surrogate models achieve 82.70\% to 83.80\% fidelity with the target hardware model, indicating strong approximation of the hardware-specific decision boundaries. Interestingly, there is minimal difference between int-8 and full-precision extraction fidelity, as Copycat CNN relies only on hard-label outputs which are preserved across quantization levels.

\subsubsection{Phase 2: Adversarial Transferability}
\label{sec:cifar10_transferability}

\begin{figure*}
     \centering
     \subfloat[][Target ResNetV1 using CIFAR-10]{\includegraphics[width=\linewidth]{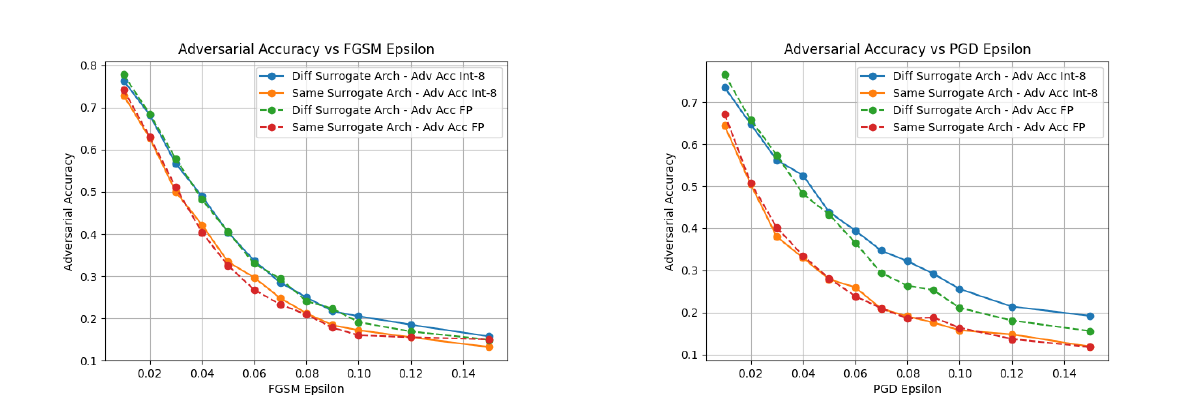}\label{fig:cifar_compare}} \\
     \subfloat[][Target MobileNetV1 0.25x using VWW]{ \includegraphics[width=\linewidth]
    {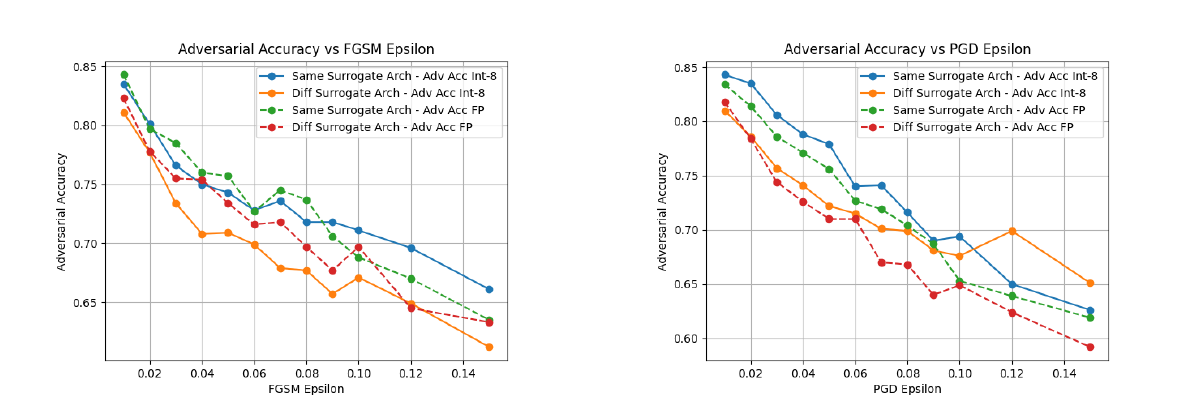} \label{fig:vww_compare}}
     \caption{Model evasion results on the target ResNetV1 using CIFAR-10 and on MobileNetV1 0.25x using VWW. The results for FGSM are on the left, and the results for PGD are on the right. Solid lines denote the adversarial accuracy of int-8 quantized models, while dotted lines denote the adversarial accuracies of full-precision models. There are separate sets of these lines for whether the same architecture of the target model was used for creating the surrogate model.}
     \label{steady_state}
\end{figure*}

Figure \ref{fig:cifar_compare} reveals a key finding: adversarial examples crafted on our surrogate models transfer equally well to both int-8 quantized hardware and full-precision models, with the solid (int-8) and dotted (full-precision) lines largely overlapping. This contradicts previous work suggesting poor transferability to quantized models \cite{costa2024david}, and indicates that our extraction process successfully captures the quantized model's behavior, enabling effective white-box attacks on hardware that would otherwise be impossible due to gradient discontinuities.

\subsubsection{Performance Against State-of-the-Art}
\label{sec:cifar10_sota}

Table \ref{tab:attack_comp} positions our approach within the landscape of existing attacks. Against int-8 hardware models, we outperform all gray-box attacks by over 20\% in accuracy reduction, likely because our surrogate-based approach circumvents the gradient masking that affects direct gray-box methods like ZOO. While black-box attacks achieve higher success rates, they require thousands of queries per image versus our one-time 50,000 query cost. Most importantly, our attack's effectiveness diminishes on full-precision models, confirming that we are exploiting vulnerabilities specific to quantized hardware implementations rather than general model weaknesses.

\begin{table*}[t]  
   \centering
   \caption{Comparison of two-step attack sequence results to direct gray-box and black-box attacks for VWW and CIFAR-10. The results for the gray-box and black-box attacks are pulled from \cite{costa2024david}, who used the same models and datasets as in our work}
   \label{tab:attack_comp}
   \begin{tabular}{@{}l | llll | llll | c@{}}  
     \toprule
     \multirow{2}{*}{\textbf{Attack}} & \multicolumn{4}{c|}{CIFAR-10}  & \multicolumn{4}{c|}{Visual Wake Words} & \multirow{2}{*}{\textbf{Work}} \\ 
      & \textbf{Accuracy} & \textbf{Adv. Acc.} & \textbf{Acc. Reduct.} & \textbf{$L_\infty$} & \textbf{Accuracy} & \textbf{Adv. Acc.} & \textbf{Acc. Reduct.} & \textbf{$L_\infty$} &  \\ 
     \midrule
     \multicolumn{10}{c}{QNN (8-bit integer)} \\
     \arrayrulecolor[gray]{0.7}\specialrule{.1em}{.05em}{.05em}
     \rowcolor[gray]{0.9}
     \multicolumn{10}{c}{\textbf{Proposal: Extraction $\rightarrow$ White-Box}} \\
     \arrayrulecolor{black}\midrule
     PGD                & 85.30\% & 38.00\% & 47.30\% & 0.03 & 86.10\% & 75.70\% & 10.40\% & 0.03 & This Work \\
     FGSM               & 85.30\% & 50.00\% & 35.30\% & 0.03 & 86.10\% & 73.40\% & 12.70\% & 0.03 & This Work \\
     \arrayrulecolor[gray]{0.7}\specialrule{.1em}{.05em}{.05em}
     \rowcolor[gray]{0.9}
     \multicolumn{10}{c}{\textbf{Gray-Box}} \\
     \arrayrulecolor{black}\midrule
     ZOO                & 87.50\% & 87.50\% & 0.00\%  & 0.00 & 91.00\% & 84.00\% & 7.00\%  & 0.09 & Costa et al. \cite{costa2024david} \\
     Square-$L_2$       & 87.50\% & 67.80\% & 19.70\% & 0.22 & 91.00\% & 64.60\% & 26.40\% & 0.32 & Costa et al. \cite{costa2024david} \\
     Square-$L_\infty$  & 87.50\% & 82.30\% & 5.20\%  & 0.01 & 91.00\% & 86.00\% & 5.00\%  & 0.01 & Costa et al. \cite{costa2024david} \\
     \arrayrulecolor[gray]{0.7}\specialrule{.1em}{.05em}{.05em}
     \rowcolor[gray]{0.9}
     \multicolumn{10}{c}{\textbf{Black-Box}} \\
     \arrayrulecolor{black}\midrule
     Boundary           & 87.50\% & 6.50\%  & 81.00\% & 0.08 & 91.00\% & 91.00\%  & 0.00\%  & 0.00 & Costa et al. \cite{costa2024david} \\
     GeoDA              & 87.50\% & 13.00\% & 74.50\% & 0.08 & 91.00\% & 91.00\%  & 0.00\%  & 0.03 & Costa et al. \cite{costa2024david} \\
    \midrule
     \multicolumn{10}{c}{DNN (32-bit floating-point)} \\
     \arrayrulecolor[gray]{0.7}\specialrule{.1em}{.05em}{.05em}
     \rowcolor[gray]{0.9}
     \multicolumn{10}{c}{\textbf{Proposal: Extraction $\rightarrow$ White-Box}} \\
     \arrayrulecolor{black}\midrule
     PGD                & 84.80\% & 40.20\% & 44.60\% & 0.03 & 86.50\% & 74.40\% & 12.10\% & 0.03 & This Work \\
     FGSM               & 84.80\% & 51.10\% & 33.70\% & 0.03 & 86.50\% & 75.50\% & 11.00\% & 0.03 & This Work \\
     \arrayrulecolor[gray]{0.7}\specialrule{.1em}{.05em}{.05em}
     \rowcolor[gray]{0.9}
     \multicolumn{10}{c}{\textbf{Gray-Box}} \\
     \arrayrulecolor{black}\midrule
     ZOO                & 88.30\% & 24.70\% & 63.60\% & 0.05 & 91.30\% & 54.00\% & 37.30\%  & 0.20 & Costa et al. \cite{costa2024david} \\
     Square-$L_2$       & 88.30\% & 47.30\% & 41.00\% & 0.22 & 91.30\% & 45.30\% & 46.00\%  & 0.28 & Costa et al. \cite{costa2024david} \\
     Square-$L_\infty$  & 88.30\% & 6.10\%  & 82.20\% & 0.01 & 91.30\% & 37.60\% & 53.70\%  & 0.01 & Costa et al. \cite{costa2024david} \\
     \arrayrulecolor[gray]{0.7}\specialrule{.1em}{.05em}{.05em}
     \rowcolor[gray]{0.9}
     \multicolumn{10}{c}{\textbf{Black-Box}} \\
     \arrayrulecolor{black}\midrule
     Boundary           & 88.30\% & 7.70\%  & 80.60\% & 0.01 & 91.30\% & 91.30\%  & 0.00\%  & 0.00 & Costa et al. \cite{costa2024david} \\
     GeoDA              & 88.30\% & 11.00\% & 77.30\% & 0.03 & 91.30\% & 91.30\%  & 0.00\%  & 0.03 & Costa et al. \cite{costa2024david} \\
     \bottomrule
   \end{tabular}
 \end{table*}

\begin{table}
    \centering
    \caption{Time spent\tablefootnote{The time has been computed multiplying the average query time of each TinyML hardware, measured over 1000 queries, by the number of queries required by each attack.} querying the target TinyML hardware by the two-step attack adopted in this paper and the Square Attack, the most query efficient alternative requiring only 1 query per iteration to generate an adversarial example. The hardware targets of this comparison are NUCLEO-F401RE (F4), NUCLEO-G474RE (G4), and NUCLEO-H743ZI2 (H7). The same setting adopted by \cite{costa2024david} have been adopted. 
    }
    \label{tab:hw_eval}
    \begin{tabular}{c | c c c | c c c}
    \toprule
    \multirow{2}{*}{\textbf{Attack}} & \multicolumn{3}{c|}{CIFAR-10} & \multicolumn{3}{c}{Visual Wake Words} \\
                      & \bf{F4} & \bf{G4} & \bf{H7} & \bf{F4}   & \bf{G4}   & \bf{H7} \\
    \midrule
    \multicolumn{7}{c}{QNN (8-bit integer)} \\
    \midrule
    Proposal          & 3.4h   & 1.8h  & 0.5h  & 2.2h  & 1.2h  & 0.5h \\ 
    Square-$L_2$      & 135.6h & 70.6h & 18.5h & 87.8h & 47.2h & 18.5h \\ 
    Square-$L_\infty$ & 67.8h  & 35.3h & 9.3h  & 43.9h & 23.6h & 9.3h \\ 
    \bottomrule
    \end{tabular}
\end{table}

\subsubsection{Feasibility on TinyML Hardware}
\label{sec:cifar10_hw_eval}
Table \ref{tab:hw_eval}, in the CIFAR-10 section, showcases the query efficiency of our approach, which translates to the fastest interaction time with the target TinyML hardware. Only 0.5 hours of querying is required with the fastest hardware, the NUCLEO-H743ZI2 (H7), while 3.4 hours is required with the slowest hardware, the NUCLEO-F401RE (F4). When compared to the most query-efficient alternative of the Square Attack, the range increases up to 67.8 hours for the slowest hardware and 18.5 hours for the fastest hardware.

\subsection{Visual Wake Words Results}
\label{sec:vww_results}

Our attack pipeline demonstrates moderate effectiveness on the Visual Wake Words hardware benchmark, achieving 12.70\% accuracy reduction with FGSM and 10.40\% with PGD on int-8 quantized models. While less dramatic than CIFAR-10, these results reveal important insights about hardware vulnerabilities in binary classification tasks. Notably, our approach outperforms some state-of-the-art gray-box attacks while maintaining the same practical advantage of requiring only 50,000 queries. The attack again shows hardware specificity, with reduced effectiveness against full-precision models.

We examine each phase to understand the factors affecting attack performance on this binary classification task.

\subsubsection{Phase 1: Model Extraction Fidelity}
\label{sec:vww_extraction}

Model extraction results for VWW (Table \ref{tab:extract_odd}, lower section) show fidelity ranging from 79.40\% to 82.10\%, slightly lower than CIFAR-10 but still indicating strong approximation of the hardware model. However, surrogate accuracy is noticeably lower than the target model accuracy (69.60\%-72.00\% vs 85.60\%), likely due to our query budget being half the original training set size (50,000 queries vs 100,000 training samples). Despite lower surrogate accuracy, the high fidelity metrics suggest the surrogate captures the hardware-specific decision boundaries necessary for adversarial generation.

\subsubsection{Phase 2: Adversarial Transferability}
\label{sec:vww_transferability}

Figure \ref{fig:vww_compare} shows more nuanced transferability patterns compared to CIFAR-10. In the FGSM plot, solid (int-8 quantized hardware) lines fall slightly below dotted (full-precision) lines, suggesting marginally better transferability to quantized hardware. Conversely, the PGD plot shows the opposite trend. These marginal differences likely stem from attack nondeterminism rather than fundamental differences between hardware and full-precision models.

The reduced attack effectiveness on VWW compared to CIFAR-10 likely stems from the fundamental geometry of binary classification. With only two classes (person vs.\ no-person), the model learns a single decision boundary rather than the complex multi-class boundaries in CIFAR-10's ten-class problem. This simpler boundary structure means perturbations must push inputs further to cross into the opposite class, and there are no "nearby" alternative classes that might capture misclassified inputs. Additionally, VWW's binary output provides less information for the surrogate to learn from during extraction, potentially resulting in coarser approximation of the true decision boundary.

\subsubsection{Performance Against State-of-the-Art}
\label{sec:vww_sota}

Table~\ref{tab:attack_comp} reveals a different competitive landscape for VWW. We acknowledge that Square-$L_2$ achieves higher accuracy reduction than our approach (26.40\% vs.\ 12.70\%). However, this comparison requires important context: Square Attack requires access to confidence scores, which real TinyML hardware rarely exposes. More critically, black-box attacks (Boundary Attack, GeoDA) achieve 0\% accuracy reduction on VWW because the binary classification boundaries defeat these methods entirely. This makes our approach the only effective option for attacking VWW on hardware that only exposes hard labels.

The full-precision results again confirm hardware specificity: our attack performs comparably poorly on both quantized and full-precision VWW models, but state-of-the-art attacks show dramatically better performance on full-precision versions. This reinforces that existing attacks fail to exploit hardware-specific vulnerabilities, while our approach, though less effective on binary classification, still targets weaknesses unique to quantized implementations.

\subsubsection{Feasibility on TinyML Hardware}
\label{sec:vww_hw_eval}
Furthermore our approach demonstrates to be better suited to interact with TinyML hardware by showcasing the fastest query time. On the slowest hardware the query time of our approach is only 2.2 hours, while the fastest query time of Square Attack on the same hardware is 67.8 hours, 30.8 times slower, according to table \ref{tab:hw_eval}, Visual Wake Words section. Notably the NUCLEO-H743ZI2 (H7) achieves the same query time as the CIFAR-10 setup, while the other targets show a faster query time. This can be explained by the 30 frames per second limit imposed by the OV7670 sensor, which acts in this case as a bottleneck for the overall system.

\subsection{The Cause Behind Hardware Specificity} \label{sec:discussion}
Previous work \cite{costa2024david} argued that the underperformance of black- and gray-box attacks in the QNN setting is caused by quantization, which may erase small perturbations, making the gradient of the attack zero, or amplify bigger perturbations, making the gradient of the attack explode. Our approach circumnavigates this effect by attacking the surrogate model, which is not quantized. However, in the DNN domain, quantization is gone, but our approach still works on the surrogate model, acting as if quantization were still there, and therefore approaching the problem in a suboptimal way, falling behind the state-of-the-art.

\section{Conclusion}
\label{sec:conclusion}
This work demonstrates the critical need for domain-specific security analysis of TinyML systems by showing how a two-step attack pipeline, designed specifically for quantized neural networks, outperforms state-of-the-art methods on TinyML hardware. By achieving high-fidelity model extraction followed by white-box adversarial generation, we circumvent the gradient discontinuities that have previously protected quantized models from gradient-based attacks.

Our approach achieves significant accuracy reduction on MLPerf Tiny benchmarks while showcasing the smallest interaction time with the TinyML target hardware. Furthermore, it is the only attack able to affect the simpler decision boundaries of the model trained on the Visual Wake Words dataset without requiring access to the class probabilities, which are rarely exposed by real-world TinyML hardware. Most importantly, the attack's effectiveness is specific to quantized implementations, confirming that we exploit hardware-induced vulnerabilities rather than general model weaknesses.

As AI accelerators proliferate across edge computing applications, ranging from autonomous systems to IoT devices, our findings emphasize that hardware-aware security must become a fundamental design consideration rather than an afterthought. Future work should focus on developing comprehensive security frameworks that account for the unique computational characteristics of quantized neural networks and the constraints of hardware deployment. Only through such domain-specific approaches can we ensure the security and reliability of the rapidly expanding TinyML ecosystem.

\bibliographystyle{IEEEtran}
\bibliography{references}

\begin{IEEEbiographynophoto}
{Jacob Huckelberry}
 received his B.Sc. in Computer Science from the United States Military Academy in 2023 and his M.Sc. in Data Science from Harvard University. Concurrently, Jacob is a Cyber Officer in the U.S. Army and a Draper Laboratory Scholar in their Cyber AI Tools group. His research interests lie at the intersection of machine learning, embedded systems, and cybersecurity. Contact him at jacobhuckelberry@g.harvard.edu.
\end{IEEEbiographynophoto}

\begin{IEEEbiographynophoto}
{Andrea Mattia Garavagno}
 received his B.Sc. degree in Electronic Engineering from the University of Genova, Italy, in 2018 and his M.Sc. degree with honours in Embedded Computing Systems, from Scuola Superiore Sant’Anna and the University of Pisa, Italy, in 2022. He's currently a Ph.D. student at the University of Genoa co-funded by Scuola Superiore Sant'Anna, Italy, and a Fellow of the Harvard John A. Paulson School of Engineering and Applied Sciences researching TinyML systems. He co-authored the book “Introduction to Microprocessor-Based Systems Design” published by Springer-Nature. Contact him at andreamattia.garavagno@edu.unige.it.
\end{IEEEbiographynophoto}

\begin{IEEEbiographynophoto}
{Yuke Zhang} 
 received the B.Eng. degree from the Beijing University of Posts and Telecommunications, Beijing, China, and the M.A.Sc. from Dalhousie University, Halifax, NS, Canada. She is currently pursuing the Ph.D. degree in electrical and computer engineering with the University of Southern California, Los Angeles, CA, USA. Her research interests include privacy-preserving machine learning and hardware security. Contact her at yukezhan@usc.edu.
\end{IEEEbiographynophoto}

\begin{IEEEbiographynophoto}
{Peter A. Beerel} 
 is the Associate Chair of Graduate Programs in the ECE Department at USC and a Principal Distinguished Scientist and Research Director at the Information Sciences Institute. He co-founded TimeLess Design Automation in 2008, which was sold to Fulcrum Microsystems in 2010 then acquired by Intel in 2011. He has (co)authored 16 US patents and more than 190 conference and journal papers covering a wide range of topics in VLSI, CAD, and Machine Learning, several of which have won best paper awards. He has been on the program committee of various VLSI conferences and has been associate editor for IEEE TCAD and IEEE Transactions on Circuits and Systems I. Contact him at pabeerel@usc.edu.
\end{IEEEbiographynophoto}

\begin{IEEEbiographynophoto}
{James Mickens}
is a Gordon McKay Professor of Computer Science at Harvard University. His research focuses on the performance and security of large-scale online services. He leverages techniques from isolated software architectures, applied cryptography, and secure hardware design. Prior to becoming a professor at Harvard, Dr. Mickens spent seven years at Microsoft Research, working in the Distributed Systems group and collaborating with various product teams from Azure and Windows; he was also a visiting professor at MIT's Parallel and Distributed Operating Systems group. At Harvard, he serves on the Board of Directors for the Berkman Klein Center for Internet \& Society. Contact him at mickens@g.harvard.edu.
\end{IEEEbiographynophoto}

\begin{IEEEbiographynophoto}
{Vijay Janapa Reddi} 
 expertise lies in computer architecture, machine learning systems, and autonomous agents. His leadership roles in MLCommons and the tinyML Foundation, along with his work in developing the MLPerf benchmarks, provide valuable insights into the practical challenges and opportunities of deploying TinyML in the real-world. Prof. Janapa Reddi’s experience in academia and industry collaborations ensures a comprehensive and practical approach to addressing the resource-constrained challenges of securing TinyML systems. Contact him at vj@eecs.harvard.edu.
\end{IEEEbiographynophoto}

\end{document}